\documentclass[a4paper, 10 pt, conference]{ieeeconf}  % Comment this line out if you need a4paper

\IEEEoverridecommandlockouts                              % This command is only needed if 
\usepackage{graphics} % for pdf, bitmapped graphics files
\usepackage{epsfig} % for postscript graphics files
\usepackage{mathptmx} % assumes new font selection scheme installed
\usepackage{times} % assumes new font selection scheme installed
\usepackage{amsmath} % assumes amsmath package installed
\usepackage{amssymb}  % assumes amsmath package installed
\usepackage[T1]{fontenc}
\usepackage{amsfonts}
\usepackage{booktabs}
\usepackage{siunitx}
\usepackage{caption} 
\usepackage{algorithm}
\usepackage{algorithmic}
\usepackage[flushleft]{threeparttable}
\usepackage[noadjust]{cite}
\usepackage{subcaption}
\usepackage{url}

\title{\LARGE \bf
Evaluating Digital Work Instructions with Augmented Reality versus Paper-based Documents for Manual, Object-Specific Repair Tasks in a Case Study with Experienced Workers$^{*}$ \thanks{$^{*}$This version of the article has been accepted for publication, after peer review but is not the Version of Record and does not reflect post-acceptance improvements, or any corrections. The Version of Record is available online at: http://dx.doi.org/10.1007/s00170-023-11313-4}
}

\author{Leon Eversberg$^{1}$, Jens Lambrecht$^{1}$% <-this % stops a space
\thanks{$^{1}$Technische Universität Berlin, Chair Industry Grade Networks and Clouds, Straße des 17. Juni 135, 10623 Berlin, Germany
        {\tt\small leon.eversberg@tu-berlin.de}}%
}

\begin{document}
\maketitle

\thispagestyle{empty}
\pagestyle{empty}

%%%%%%%%%%%%%%%%%%%%%%%%%%%%%%%%%%%%%%%%%%%%%%%%%%%%%%%%%%%%%%%%%%%%%%%%%%%%%%%%
\begin{abstract}
Manual repair tasks in the industry of maintenance, repair, and overhaul require experience and object-specific information. Today, many of these repair tasks are still performed and documented with inefficient paper documents. Cognitive assistance systems have the potential to reduce costs, errors, and mental workload by providing all required information digitally. In this case study, we present an assistance system for object-specific repair tasks for turbine blades. The assistance system provides digital work instructions and uses augmented reality to display spatial information. In a user study with ten experienced metalworkers performing a familiar repair task, we compare time to task completion, subjective workload, and system usability of the new assistance system to their established paper-based workflow. All participants stated that they preferred the assistance system over the paper documents. The results of the study show that the manual repair task can be completed 21~\% faster and with a 26~\% lower perceived workload using the assistance system.
\end{abstract}

\vspace{5px}

%%%%%%%%%%%%%%%%%%%%%%%%%%%%%%%%%%%%%%%%%%%%%%%%%%%%%%%%%%%%%%%%%%%%%%%%%%%%%%%%
\section{Introduction}
In contrast to the paradigm of workerless factories from the 1980s, humans remain the most flexible entities in the era of Industry 4.0 and Smart Manufacturing~\cite{Gorecky2014}.
However, complexity is ever-increasing due to new manufacturing paradigms such as shorter product life cycles, increasing product variety, and mass customization~\cite{Brinzer2020,Alkan2018}.
To overcome these challenges, industrial assistance systems can be utilized in order to enhance a person's physical, sensorial or cognitive capabilities, leading to a so-called Operator~4.0~\cite{Romero2016}. Cognitive assistance systems~(CAS) enhance cognitive capabilities by providing digital work instructions, helping users make decisions, and assisting them in learning new tasks~\cite{Mark2021d}.
The digitization of paper-based documents can lead to significant increases in efficiency, as providing digital work instructions removes the need for printing, searching, filing, and retrieving paper-based documents~\cite{Heng2019}.

With traditional paper-based work instructions, e.g., for assembly or maintenance tasks, the user must first search for relevant documents. Then, he has to switch between the two-dimensional work instructions, which tell him how to perform the task, and the execution of the task on the three-dimensional physical object~\cite{Jeffri2021}. 
According to Cognitive Load Theory, splitting the user's attention between spatially or temporally separated information sources leads to additional cognitive load~\cite{Sweller2011a}. This so-called Split-Attention Effect can be reduced with Augmented Reality (AR)~\cite{Dixon2018}, which superimposes virtual objects onto the real world.

Research on industrial CAS with AR has seen a steady increase in the number of publications over the last decade and focuses primarily on manual assembly and maintenance tasks~\cite{Bottani2019Augmented, Egger2020Augmented, SouzaCardoso2020}. For those applications, digital work instructions can be semi-automatically generated, given that each task follows a standardized meta-model of a workflow~\cite{Lindorfer2018, Fabian2016, Geng2020systematic}.

Manual repair tasks in the industry of maintenance, repair, and overhaul~(MRO) are less standardized and instead require intuition- and experienced-based decision-making from experienced workers \cite{Esposito2019, Bertram2020}, e.g., for the repair of turbine blades or fiber-reinforced composite structures in an aircraft.
In our previous work~\cite{Eversberg2022ML}, we described the system architecture of a CAS based on the digital twin for manual, object-specific repair tasks.
In this present work, we describe the evaluation of our CAS in a case study with experienced workers on a familiar turbine blade repair task.
The main contributions of this work are the following:
\begin{itemize}
    \item the presentation of a CAS for manual, object-specific repair tasks which are frequently performed in the MRO industry, and
    \item experimental results from a user study with experienced shop floor metalworkers comparing the presented CAS to their current paper-based workflow.
\end{itemize}

\section{Related Works}
\label{ch:related_works}
Research on CAS with AR has been focusing primarily on manual assembly and maintenance tasks~\cite{Bottani2019Augmented,Egger2020Augmented, SouzaCardoso2020}.
Multiple studies have shown the benefits of CAS with AR, such as reducing the time to complete manual tasks~\cite{Illing2020, Ababsa2020}, human errors~\cite{Funk2016Interactive, Chu2020Comparing}, mental workload~\cite{Jeffri2021}, or improving the learning curve for new tasks~\cite{Hou2013, Kaestner2021}.
The three most commonly used metrics for evaluating industrial assistance systems are task completion time~(TCT), the number of human errors in the process, and the subjectively perceived workload with the NASA-TLX survey~\cite{Egger2020Augmented, Jeffri2021}. 

%manual assembly
Funk et al.~\cite{Funk2017Teach} developed an assistance system for manual assembly equipped with a Microsoft Kinect 3D camera and a projector for AR work instructions. Their assistance system recognizes the current assembly step by evaluating pick locations, the current assembly workpiece, and a predefined tool zone. In an 11-day user study at an assembly line of a car manufacturing company, they found that the in-situ projections were useful during the learning phase of untrained workers. However, for expert workers, the assistance system had a negative effect, i.e., perceived workload and TCT increased~\cite{Funk2017Working}.

In \cite{Lai2020Smart}, Lai et al. developed a CAS for manual assembly tasks. They equipped a workbench with two webcams and a display for AR. A neural network was trained on synthetic images to detect relevant tools with a webcam and highlight their position in digital work instructions. In a user study with 20 university students with no prior experience, participants were asked to carry out a spindle motor assembly task. Compared to paper-based documents, the CAS reduced the TCT by 33~\% and the number of errors during the task by 32~\%.

Uva et al.~\cite{Uva2016Design} developed a projective AR workbench for assembly and maintenance tasks. A projector was used to visualize text and 2D symbols on the workbench and on the maintenance object. The maintenance object was mounted on a movable tracking board with markers. A user study was conducted in \cite{Uva2017Evaluating} with 16 untrained engineering students to compare projective AR with paper-based documents. For a combination of assembly and disassembly tasks on a motorbike engine, they found a 20~\% reduction in TCT using the assistance system, as well as an improvement in error rate and subjectively better ease of use, satisfaction level, and intuitiveness. In conclusion, they point out that the assistance system is particularly beneficial for tasks with high complexity.

Kästner et al.~\cite{Kaestner2021} presented an assistance system with monitor-based AR for complex manual tasks. They used deep learning models for object detection and action recognition to automatically detect the current work step and display corresponding work instructions. In a user study with 30 inexperienced participants, they found that the group with AR assistance had a lower TCT than the group without AR assistance in the first 13 iterations. However, for more than 13 iterations, the group without AR assistance performed better than the AR group. Therefore, it was concluded that AR assistance is beneficial for untrained workers in the first iterations when learning a new task.

Havard et al.~\cite{Havard2021} conducted a user study with 20 engineering students to compare PDF maintenance instructions with AR maintenance instructions on a mobile tablet. The maintenance process consisted of 27 actions to replace two springs in a machine.
Several markers were placed on the machine for spatial registration of the AR work instructions. 
In the evaluation, no statistically significant difference was found for either group in terms of TCT or mental workload using NASA-TLX.
In summary, they recommend AR work instructions for training inexperienced workers in sufficiently complex tasks and tasks that are frequently performed.

In \cite{Vanneste2020}, AR work instructions using a projector were compared to paper and oral work instructions for three different assembly tasks. 44 participants, some of whom were cognitively or physically impaired, participated in the user study. The evaluation found that with AR work instructions, participants had a lower perceived task complexity and made fewer errors. However, there was no significant difference in TCT.
Vanneste et al. point out that the advantages of AR diminish over time with repetition, therefore AR should be used for less experienced workers.

In summary, current research on CAS in manufacturing has focused primarily on manual assembly and maintenance. In the application area of maintenance, research on assistance systems has been focusing on standardized replacement repair according to the scheme of step-by-step disassembly, component replacement, and step-by-step assembly~\cite{Fiorentino2014,Henderson2011,Uva2017Evaluating,Havard2021,Obermair2020Maintenance}.
Evaluations of industrial assistance systems mostly use inexperienced participants in their user study, e.g., computer science students, who are not familiar with the evaluated task~\cite{Merino2020}. Evaluation results often show that positive effects from the usage of CAS diminish over time, thus most industrial assistance systems are found to be useful for training new employees but not necessarily for daily use by experienced workers.

\section{Research Questions}
In order to be used in daily productive operations by experienced workers, a new assistance system must provide a noticeable improvement over the established workflow.
Based on the literature from Section~\ref{ch:related_works}, there is a gap in research regarding the evaluation of CAS with experienced workers on familiar real-world tasks.

In \cite{Eversberg2022ML} we described the development of a CAS for the manual repair process of turbine blades in MRO. The system was developed with a human-centered design approach specifically for experienced workers.
To evaluate the developed assistance system, we formulate the following two research questions (RQ) for this case study:

\begin{itemize}
    \item RQ 1: \textit{What effect does the developed CAS have on the TCT of experienced workers when performing familiar repair tasks on turbine blades?}
    \item RQ 2: \textit{What effect does the developed CAS have on the perceived workload of experienced workers when performing familiar repair tasks on turbine blades?}
\end{itemize}

\section{Material and methods}
The system architecture of the developed CAS has been described in our previous work in \cite{Eversberg2022ML}. Thus, only a brief overview of the system and its assistance functions will be given in Subsection~\ref{ch:cas}. Afterward, the conducted user study will be described in detail in Subsection~\ref{ch:user_study}.

\subsection{The Cognitive Assistance System}
\label{ch:cas}
\begin{figure}[htb]
    \centering
    \includegraphics[width=0.49\textwidth]{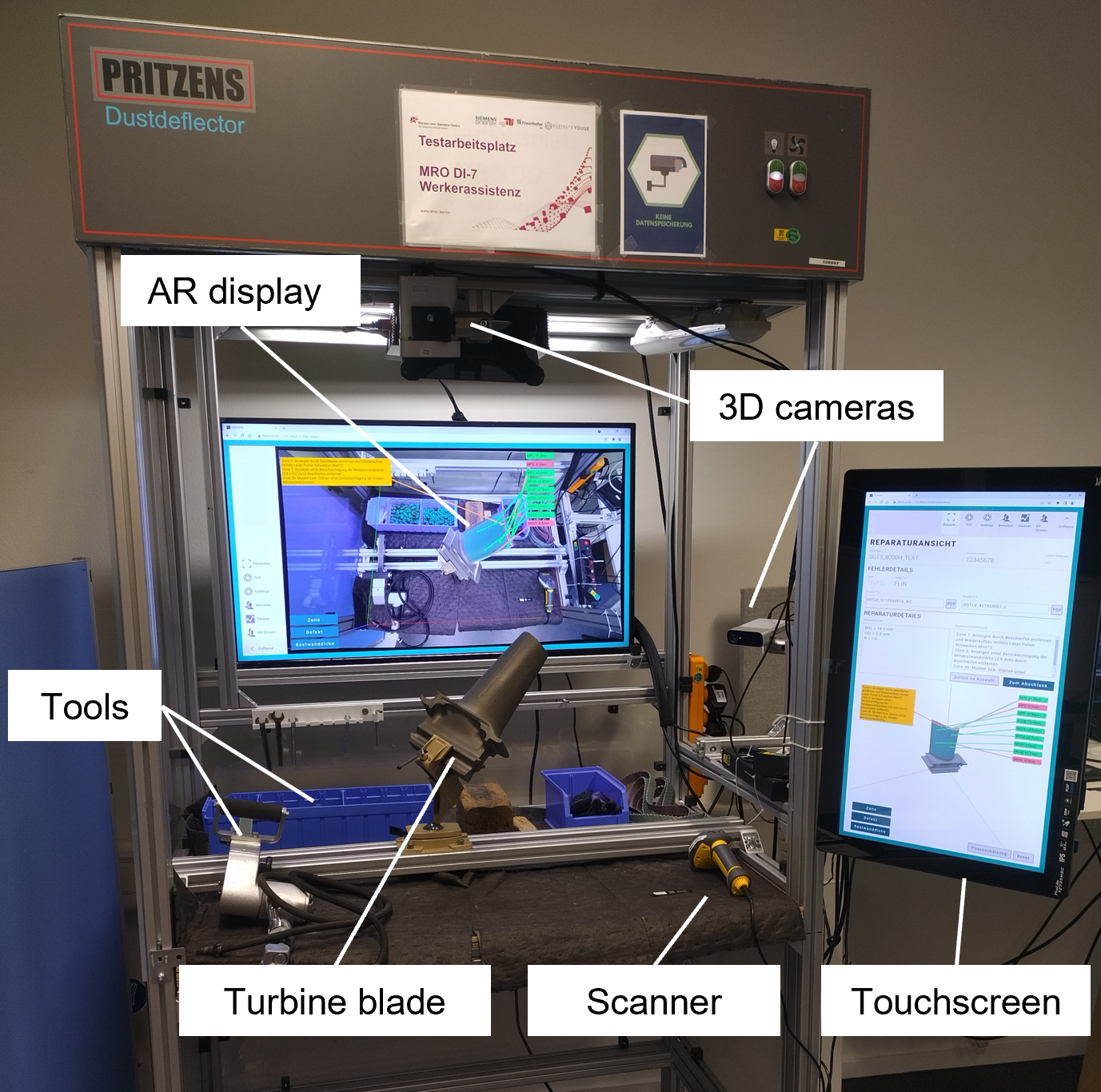}
    \caption{Our CAS supports shop floor metalworkers during the manual grinding of turbine blades. Based on the scanned serial number of a turbine blade, digital work instructions are presented on the touchscreen and on the AR display.}
    \label{fig:assistance_system}
\end{figure}

As shown in Fig.~\ref{fig:assistance_system}, we equipped a manual workstation for grinding work on turbine blades with the following additional hardware:

\begin{itemize}
    \item two Microsoft Azure Kinect 3D cameras for context awareness,
    \item a Cognex DataMan DM 8600 scanner to read 1D and 2D bar codes,
    \item a 27-inch touchscreen monitor for user interaction and
    \item a 43-inch monitor to show screen-based AR content.
\end{itemize}

\begin{figure}[htb]
\centering
	\includegraphics[width=\linewidth]{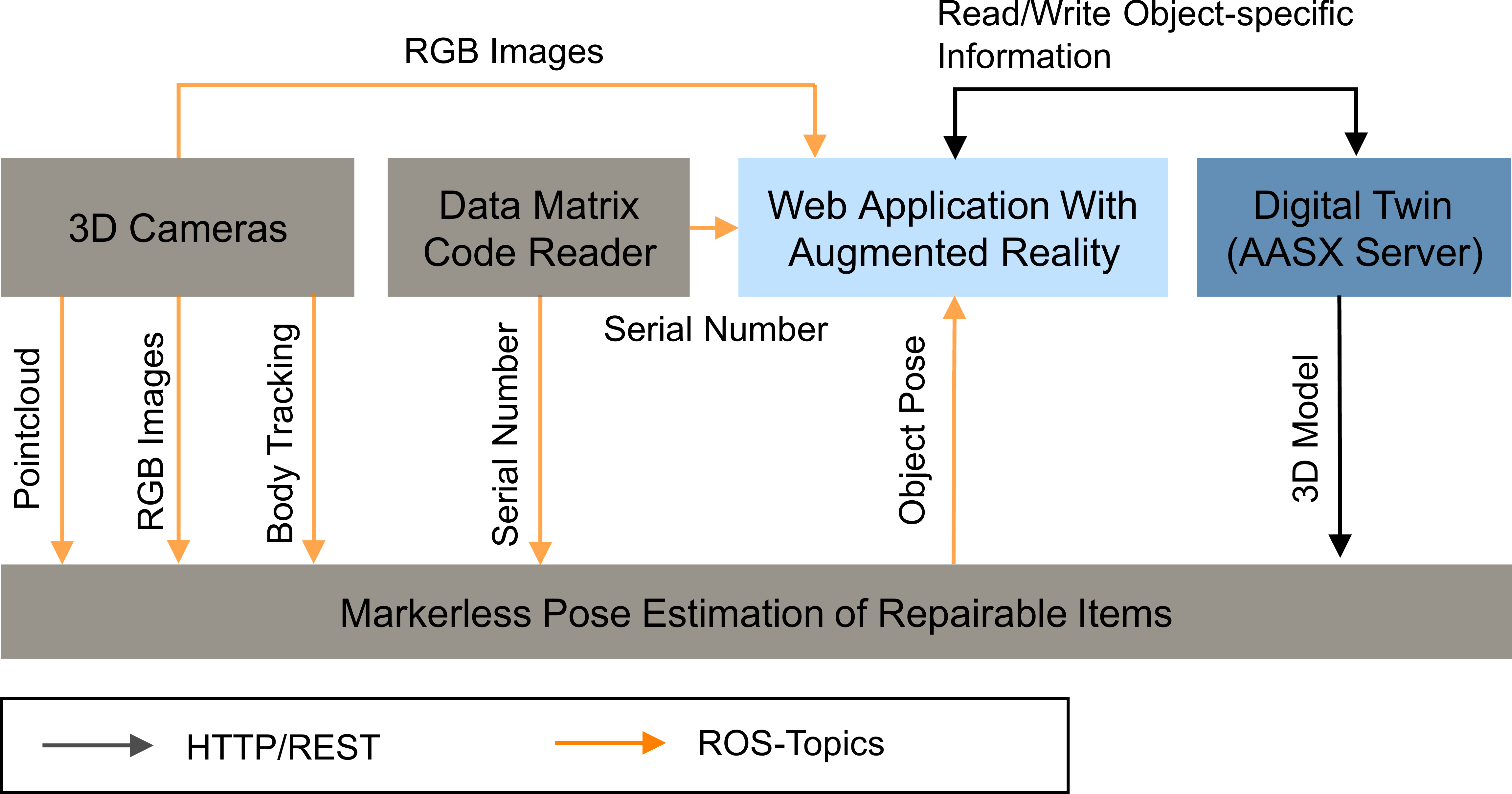}
	\caption{Overview of the system components and their communication}
\label{fig:architecture}
\end{figure}

We chose the Robot Operating System (ROS) as a framework for asynchronous many-to-many communication between our software modules through ROS topics. The system's components, depicted in Fig.~\ref{fig:architecture}, can be grouped into ROS nodes, a dynamic web application as a human-machine interface, and a digital twin for object-specific data. 

We use the Asset Administration Shell (AAS)~\cite{Neidig2022} to represent digital industrial assets. The AAS metamodel is documented in \cite{Part1AAS2020} and the HTTP/REST API is documented in \cite{Part2AAS2020}. Digital twins according to the AAS specifications can be created using available open-source AASX tools from the Industrial Digital Twin Association, i.e., the AASX Package Explorer and the AASX Server\footnote{\url{https://github.com/admin-shell-io} [accessed: 15-03-2023]}. 
Given an object's serial number, object-specific information can be accessed via the AASX server's REST API using the HTTP GET method and updated using the HTTP PUT method. 
Based on the data from the AASX Server, we implemented the following assistance functionalities for this case study:

\subsubsection{Automatic Part Identification}
Currently, the shop floor workers must manually read the serial number on each turbine blade and then look it up in a paper file, which is slow and prone to error. Therefore, we added a barcode scanner for automatic part identification. 
Specifically, we use 2D Data Matrix codes~\cite{ISO16022DataMatrix} to encode the serial number, which is applied onto each turbine blade via direct part marking. Each time a Data Matrix code is scanned, its serial number is published to the assistance system in a ROS topic. The serial number is then used to access data from the AASX Server.

\subsubsection{Digital Work Instructions}
We developed a web application to display digital work instructions on the touchscreen.
In general, the web application first provides an overview of the open tasks for the given serial number, then detailed information about the selected task, and the ability to provide digital documentation at the end of the task. The developed web application is depicted in Fig.~\ref{img:CAS_appendix} in the appendix.

After scanning a serial number, defects for the current object are shown in a list. The task of each metalworker in this case study is to repair all defects classified as open.
All instructions are linked to the serial number of the current object and are dynamically requested using the REST API of the AASX server. The digital work instructions include the following core features:

\begin{itemize}
    \item an interactive 3D model of the current object with spatial information about defects, zones, and residual wall thickness measurements,
    \item detailed text information about the selected defect, such as its type, length, and additional comments,
    \item opening PDF documents linked to the serial number,
    \item a glossary for common abbreviations, and
    \item digital documentation of the performed repair task.
\end{itemize}

\subsubsection{Screen-Based Augmented Reality}
According to the pinhole camera model, the relationship between a 3D point $\tilde{\mathbf{M}} = [X, Y, Z, 1]^T$ and its 2D image projection  $\tilde{\mathbf{m}} = [u, v, 1]^T$ is given by Eq.~\ref{eq:projection}, where $s$ is an arbitrary scale factor, $\mathbf{A}$ is the camera intrinsic matrix and $(\mathbf{R},~\mathbf{t})$ is the extrinsic rotation and translation between world coordinate system and camera coordinate system~\cite{Zhang2000}.

\begin{equation}
    s \tilde{\mathbf{m}} = \mathbf{A} \left[ \mathbf{R} ~ \mathbf{t} \right] \tilde{\mathbf{M}} \label{eq:projection}
\end{equation}

\begin{figure*}[hbt]
         \includegraphics[width=\textwidth]{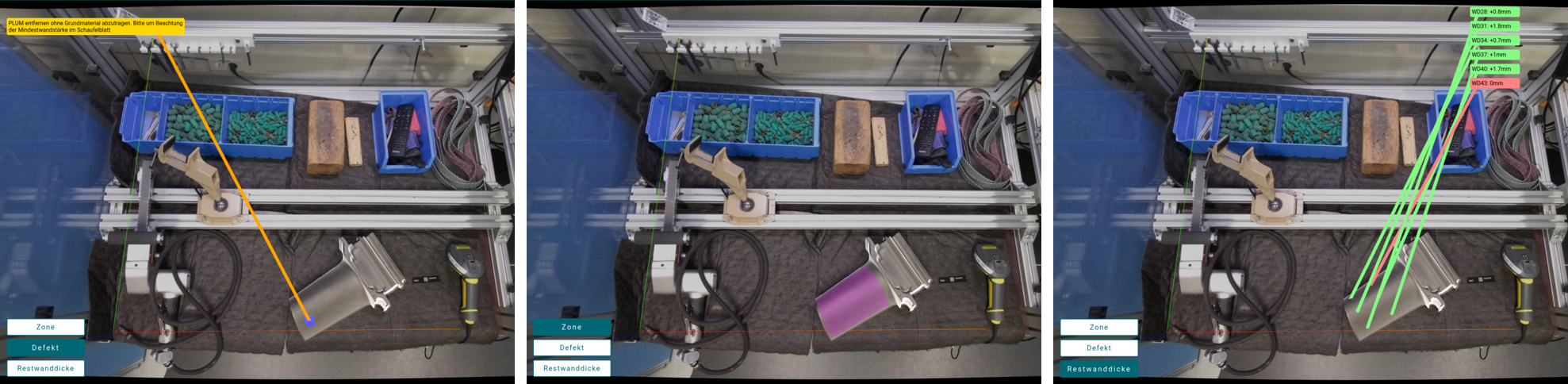}
         \caption{Screen-based AR superimposing a defect (left), its zone (middle) and nearby wall thickness measurements (right) on the live camera image}
         \label{fig:AR_details}
\end{figure*}

Based on Eq.~\ref{eq:projection}, we use the top-mounted RGB camera for screen-based AR by projecting three-dimensional AR objects into the live camera image. In general, we use AR to help the worker locate spatial information about the object being repaired.
For the manual repair of turbine blades, we superimpose defects, zones, and residual wall thickness measurements on the live image of the work area, see Fig.~\ref{fig:AR_details}. The user can toggle which information is displayed.

%NEW AR details

Each time the object being repaired is moved, we have to recompute the extrinsic parameters~$(\mathbf{R},~\mathbf{t})$. This problem is known as object pose estimation. Using homogeneous transforms~$\mathbf{T}$, we can rewrite Eq.~\ref{eq:projection} to Eq.~\ref{eq:AR} with a dynamic transformation $\mathbf{T}_O^W$ from object coordinates to world coordinates and a static transformation~$\mathbf{T}_W^{C_1}$ from world coordinates to the top-mounted camera coordinates. The camera intrinsic matrix~$\mathbf{A}$ and the transformation~$\mathbf{T}_{W}^{C_1}$ can be obtained using standard camera calibration techniques~\cite{Zhang2000}.

\begin{equation}
   s \tilde{\mathbf{m}} = \mathbf{A}
    \begin{bmatrix}
  1 & 0 & 0 & 0 \\ 
  0 & 1 & 0 & 0 \\ 
  0 & 0 & 1 & 0
  \end{bmatrix}
    \mathbf{T}_{W}^{C_1} ~ \mathbf{T}_{O}^{W} ~ \tilde{\mathbf{M}}
  \label{eq:AR}
\end{equation}

\begin{figure}[htb]
\centering
	\includegraphics[width=\linewidth]{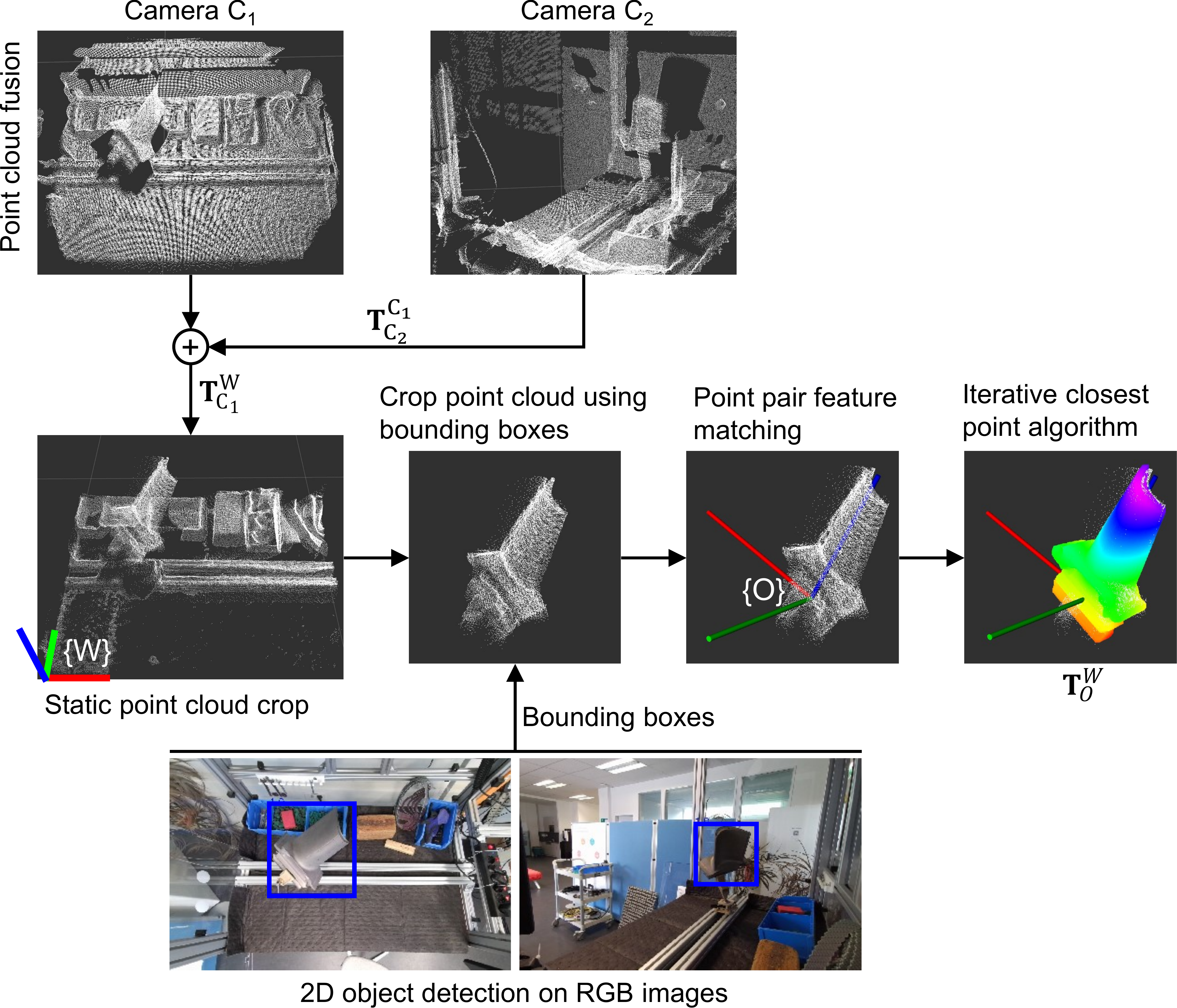}
	\caption{Overview of the pose estimation process}
\label{fig:pose_estimation}
\end{figure}

To obtain the transformation~$\mathbf{T}_O^W$, we use the markerless pose estimation process depicted in Fig.~\ref{fig:pose_estimation}.
A new pose estimation is triggered by the user's hands leaving the work area. Then, the object is first localized using 2D object detection on RGB images~\cite{EversbergSensors2021}. The obtained bounding boxes are used to crop the point cloud data from both 3D cameras. Based on the cropped point cloud, the 6D object pose $\mathbf{T}_{O}^{W}$ is estimated using point pair feature matching~\cite{Drost2010} and then refined using the point-to-plane iterative closest point algorithm~\cite{Chen1992}.

\subsection{User Study}
\label{ch:user_study}
\subsubsection{Participants}
In order to perform the evaluated manual repair task, study participants had to be professionals with a certain level of work experience.
Therefore, a total of ten male metalworkers with a median age of 46 years, ranging from 33 to 57 years, were recruited to participate in the user study. All participants were experienced in the grinding repair process of turbine blades, with a median work experience of 14 years, ranging from a minimum of 3 years to a maximum of 28 years. 

\subsubsection{Study Design}
In a between-subjects experimental study design, each participant is assigned to a single group, such as a control group or the CAS group. In a within-subjects study design, participants are assigned to multiple groups. Because within-subjects study designs usually have greater statistical power, they require fewer participants~\cite{Keren1993}.

Because of the limited number of study participants, a within-subjects study design was performed, i.e., each participant used both systems~(paper file and assistance system). To reduce practice effects, the sequence of the two systems was counterbalanced. That is, the first participant started with the CAS and then performed the same task with the paper file. The second participant started with the paper file and then used the CAS.

In the first phase, all participants watched a four-minute-long introduction video explaining all the relevant features of the CAS. Afterward, each participant had up to ten minutes to get used to working with the assistance system. Next, all participants were given up to ten minutes to check the paper-based work instructions contained in a file. Because the file was designed to replicate their current work instructions, most participants did not require much time. During this phase, the CAS and the paper-based file did not include the relevant defects of the evaluation.

In the second phase, each participant was asked to complete the given task according to Subsection~\ref{ch:task} consecutively with both systems. Because the given repair task requires mostly cognitive reasoning, and it was not feasible to physically repair real defects, participants were asked to realistically indicate the physical grinding process with unplugged tools. During the task, we measured the TCT for each turbine blade as a dependent variable. This phase was recorded on video with sound for further evaluation.

In the final phase, each participant was asked to fill out standardized questionnaires for each system and provide feedback on the CAS. We used the NASA-TLX questionnaire~\cite{Hart1988Development} to measure cognitive load and the UMUX-LITE questionnaire~\cite{Lewis2013} to measure system usability. 

\subsubsection{Task}
\label{ch:task}
The aim of the given task was to simulate a familiar real-world repair job of multiple turbine blades. To this end, we visited the shop floor and reviewed the metalworker's labor practices and required documents. 

\begin{figure*}[htb]
     \centering
     \begin{subfigure}[b]{0.49\textwidth}
         \centering
         \includegraphics[width=\textwidth]{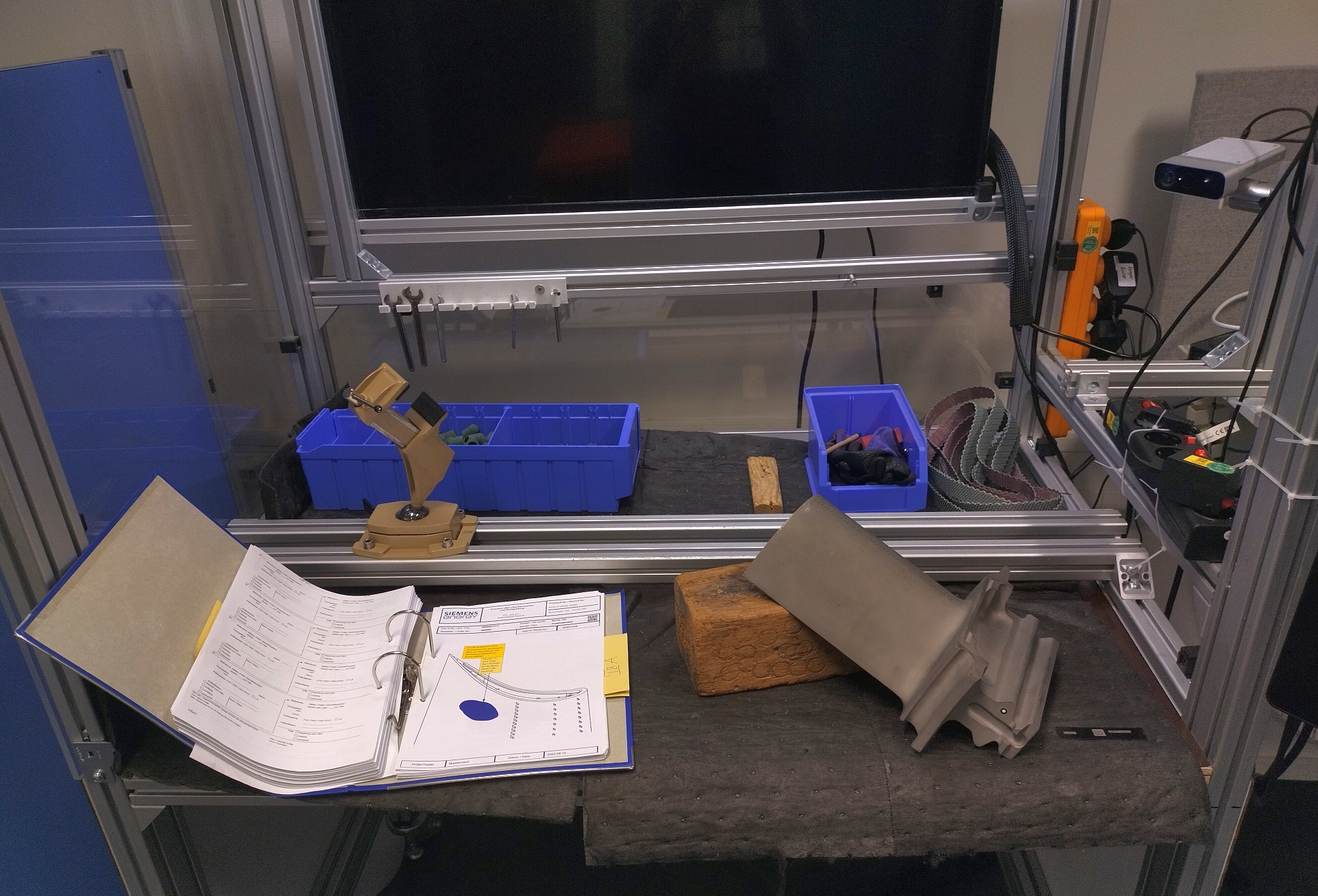}
         \caption{Set-up with the well-known paper file}
         \label{fig:paper_task}
     \end{subfigure}
     \hfill
     \begin{subfigure}[b]{0.49\textwidth}
         \centering
         \includegraphics[width=\textwidth]{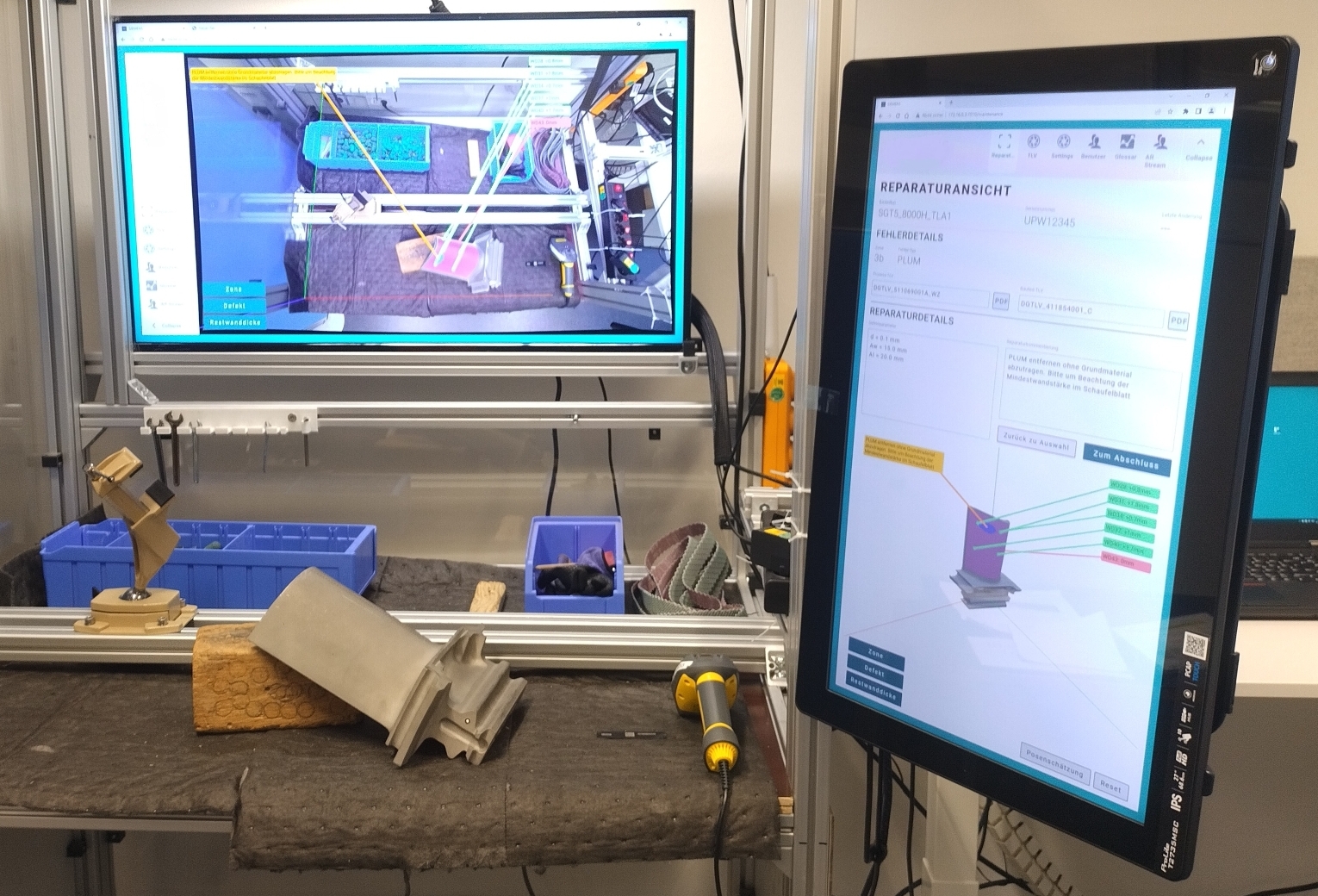}
         \caption{Set-up with the new digital assistance system}
         \label{fig:CAS_task}
     \end{subfigure}
        \caption{Study set-up for both systems}
        \label{img:task_setup}
\end{figure*}

For this study, participants were given three different serial numbers with one defect each. The participants were asked to repair any assigned defects on the surface of the three corresponding turbine blades. The necessary amount of material that has to be removed depends on the defect type, its assigned zone, and nearby residual wall thickness measurements. Additionally, they were asked to document their work and the required amount of time. All required information was either contained in a paper file or provided digitally by the assistance system. The study set-up for the paper file is depicted in Fig.~\ref{fig:paper_task} and the CAS is shown in Fig.~\ref{fig:CAS_task}.
For a more detailed workflow for both systems, see Figs.~\ref{img:paper_appendix} and \ref{img:CAS_appendix} in the appendix.

\begin{figure}[htb]
    \centering
    \includegraphics[width=0.4\textwidth]{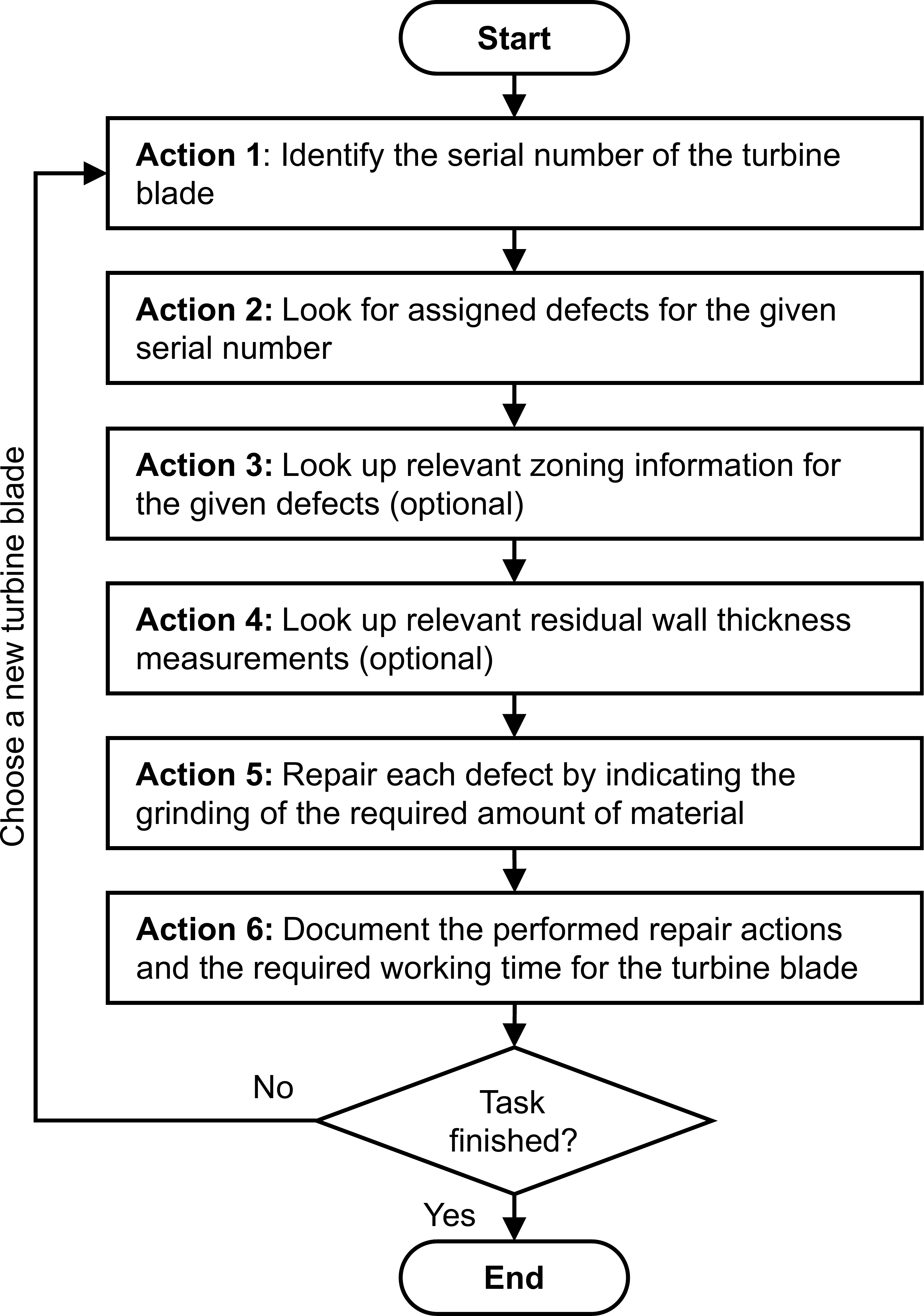}
    \caption{Observed workflow of the given task for three turbine blades}
    \label{fig:workflow}
\end{figure}

To be consistent with the participants' usual work routine, we did not impose step-by-step work instructions for the task. To complete the given task, all workers performed the actions depicted in Fig.~\ref{fig:workflow} based on their work experience. 

\section{Results}
\label{ch:results}
\subsection{Task Completion Time}
Overall, 7 out of 10 participants completed the task faster with the CAS compared to the paper file.
The time of the paper-based workflow was strongly dependent on whether action~3 and action~4 were carried out or skipped. Additionally, we observed that finding the defect information page in the paper file (action~2) often took a long time.

For the CAS, time was often lost due to the unfamiliar navigation of the digital work instructions. Furthermore, after moving the turbine blade, participants had to wait each time roughly one second for the pose estimation of the AR.

\begin{figure}[htb]
    \centering
    \includegraphics[width=0.49\textwidth]{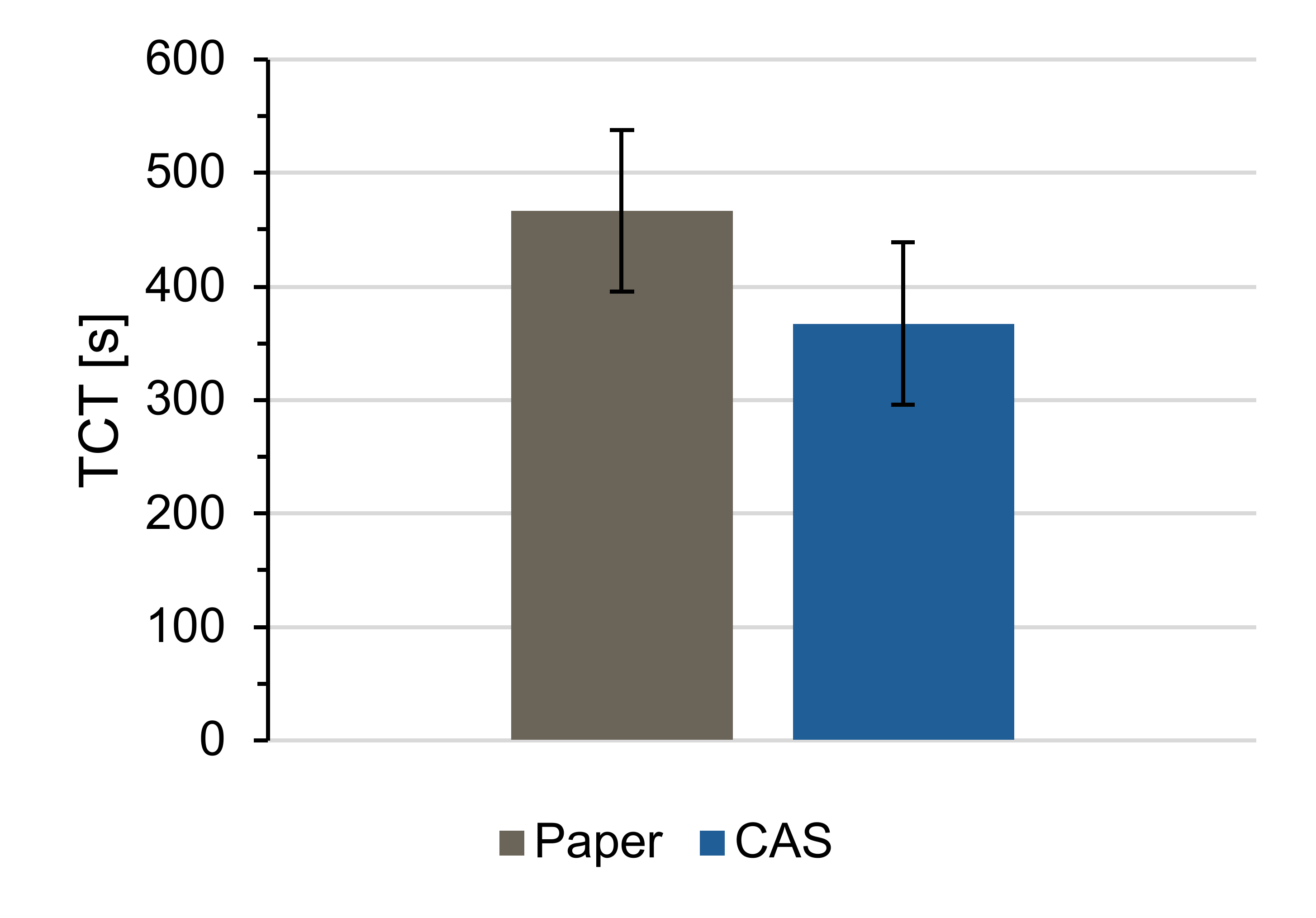}
    \caption[Average TCT for both systems with 95~\%~CI]{Average TCT for both systems with 95~\%~CI\footnotemark}
    \label{fig:tct}
\end{figure}

The mean results for TCT are shown in Fig.~\ref{fig:tct} with 95~\%~confidence intervals (CI). With the CAS, participants were on average 21.2~\% faster than the participants with the paper file.

\footnotetext{Within-subject confidence intervals were calculated according to \cite{Cousineau2014}}

We performed a paired t-test to compare the overall TCT for both systems (significance level $\alpha = 0.05$).
The test's normality assumption was checked with the Shapiro-Wilk test.
Results of the two-tailed paired t-test indicate that there is a non-significant medium difference between the paper file ($M = 466.1$, $SD = 171.3$), and the CAS ($M = 367.3$, $SD = 95.5$), $t(9) = 2.2$, $p = 0.054$.
According to Cohen's $d_z$, which measures the standardized mean difference~\cite{Cohen1988}, the effect size of the CAS on the TCT was medium with $d_z = 0.7$.

\subsection{Perceived Workload}
The perceived workload was measured with the NASA Task Load Index questionnaire~\cite{Hart1988Development}. A meta-analysis of 556 studies \cite{Hertzum2021} found that the average score for this questionnaire is at RTLX~=~42.

\begin{figure}[htb]
    \centering
    \includegraphics[width=0.49\textwidth]{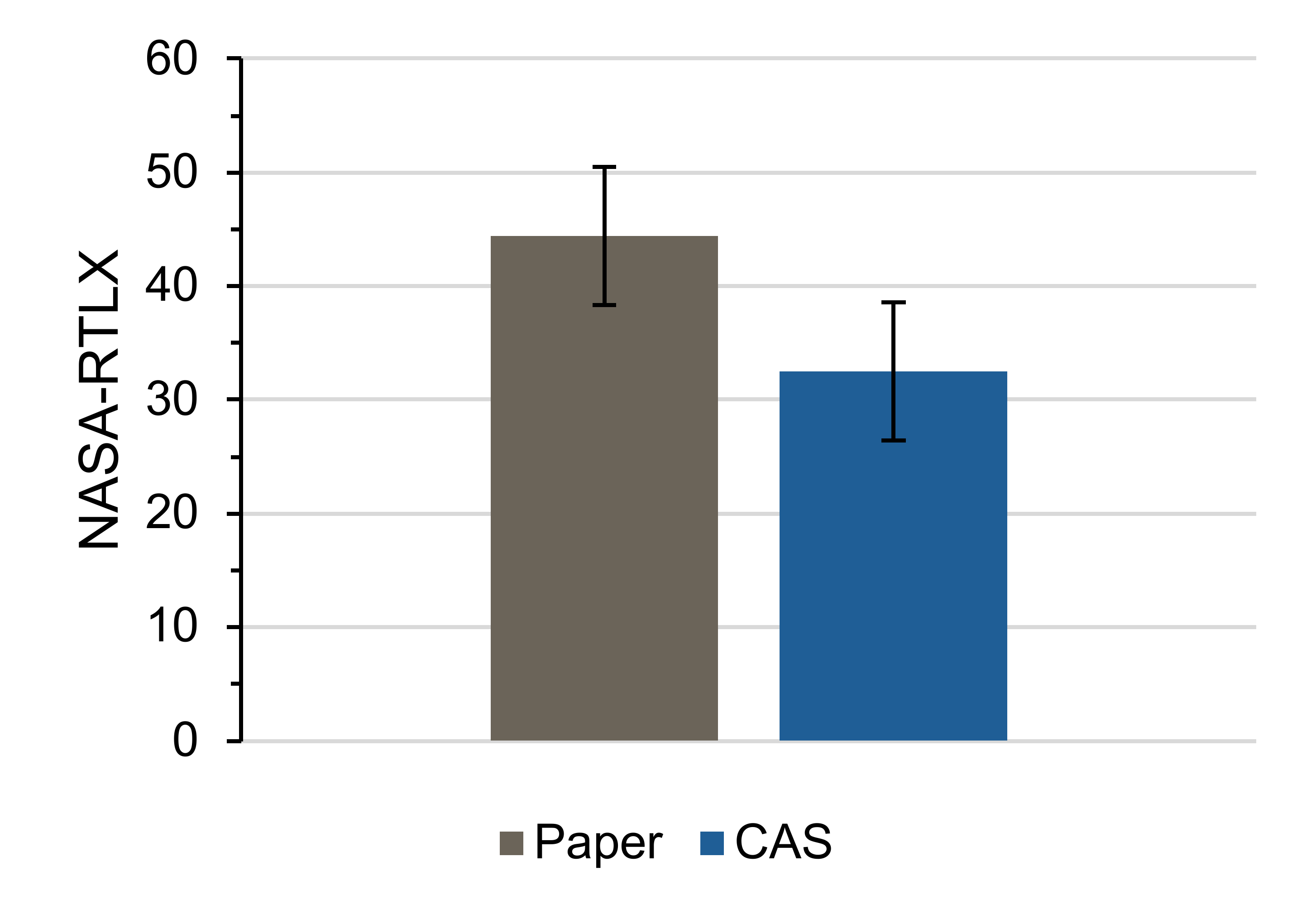}
    \caption[Perceived workload according to NASA-RTLX for both systems with 95~\% CI]{Perceived workload according to NASA-RTLX for both systems with 95~\% CI\footnotemark[2]}
    \label{fig:nasa_rtlx}
\end{figure}

Fig.~\ref{fig:nasa_rtlx} shows the RTLX values for both systems. Using the CAS, participants reported on average a 26.8~\% reduction in perceived workload compared to the paper-based workflow. 
We statistically compared RTLX values for both systems with a paired t-test ($\alpha = 0.05$). The test's normality assumption was checked with the Shapiro-Wilk test. Results of the two-tailed paired t-test indicate that there is a significant large difference between the paper file ($M = 44.4$, $SD = 16.2$) and the CAS ($M = 32.5$, $SD = 12.6$), $t(9) = 3.1$, $p = 0.012$. The effect size of the CAS on the perceived workload was large with Cohen's $d_z = 0.9$.

\subsection{System Usability}
To rate the usability of both systems, we used the UMUX-LITE~\cite{Lewis2013} questionnaire. UMUX-LITE is a short questionnaire that rates the perceived usefulness~(PU) and the perceived ease of use (PEU) of a system on a seven-point Likert scale and then combines them to a score ranging from 0 to 100. Table~\ref{table_UNUX} shows the UMUX-LITE scores and the corresponding school grades on the Sauro/Lewis curved grading scale (CGS)~\cite{Lewis2018}.

The UMUX-LITE shows poor ratings for the paper-based workflow, both for PU and PEU. The CAS was perceived as both very useful and easy to use.
During the feedback, all ten participants stated that they would prefer to use the CAS over the paper file in production.

\begin{table}[!htb]
\centering
\renewcommand{\arraystretch}{1.3}
\caption{Average UMUX-LITE scores for the usability of both systems}
\begin{tabular}{ccccc}
\hline
  & PU & PEU & UMUX-LITE & CGS Grade \\ 
\hline
Paper & 3.9 & 4.8 & 55.8 & D \\ 
\hline
CAS & 6.2 & 6.1 & 85.8 & A+ \\
\hline
\hline
\end{tabular}
\label{table_UNUX}
\end{table}

\subsection{Ranking of Individual Assistance Functions}
\begin{figure}[htb]
    \centering
    \includegraphics[width=0.49\textwidth]{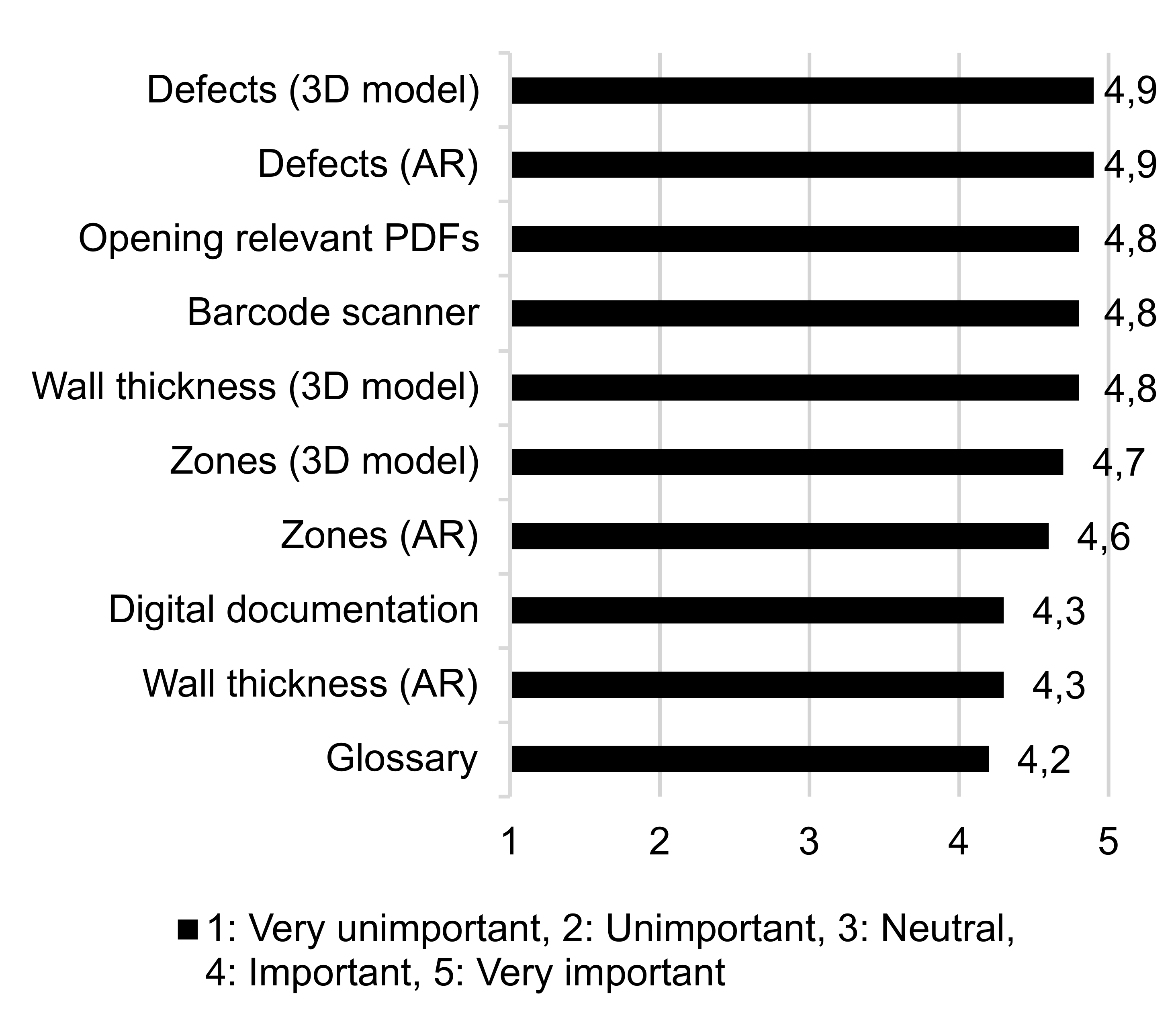}
    \caption{Average ranking of individual assistance functions on a five-point Likert scale}
    \label{fig:likert}
\end{figure}

We asked all participants to rate each implemented assistance function according to how important it was to them. The average results from the five-point Likert scale are presented in Fig.~\ref{fig:likert}. All assistance functions were rated above important. The most important function was to see the defects on the touchscreen and on the AR display. Although the glossary was not used by any participants during the task, they rated the feature as nice to have and important for new employees.

\section{Discussion}
The results from Section~\ref{ch:results} showed better efficiency, perceived workload, and usability with the CAS, compared to the paper file. 
However, results from the user study are limited to experienced male metalworkers with a median age of 46. We did not test the CAS on workers with less than three years of work experience. 

Regarding RQ~1, the TCT was reduced by an average of 21.2~\% for a familiar turbine blade repair task by experienced workers using the CAS. The paired t-test showed a non-significant medium effect for the time difference between the CAS and the paper file. 
We observed that some participants skipped action~3 and action~4 for the paper-based workflow, i.e., they did not check zoning information and residual wall thickness measurements in the paper file (see Figs.~\ref{fig:paper_action3}, \ref{fig:paper_action4_points}, and \ref{fig:paper_action4_table}). Action~3 can be skipped with enough work experience, but neglecting the measurements from action 4 can potentially lead to errors. In comparison, the CAS always displayed all required information by default. Therefore, we presume that the CAS will reduce manual errors in the long term.

The group of participants had a median work experience of 14 years repairing turbine blades with the paper file, whereas the time to practice the usage of the CAS was limited to a maximum of ten minutes during the user study. We observed that some participants lost time during the task due to the unfamiliar navigation of the digital work instructions. To this effect, multiple participants stated that if they had more time to get used to working with the assistance system, the CAS would be much faster than their current paper-based workflow. 

Because all participants had to have a certain level of work experience in the evaluated manual repair task, we were unable to recruit more than ten study participants. The small sample size of ten resulted in a low statistical power of the conducted t-tests. The statistical power is the probability to reject a null hypothesis, i.e., the probability to find a statistically significant difference in the observed groups~\cite{Cohen1988}. Statistically significant results for the TCT might be found by repeating the study with a bigger sample size, or by giving participants more time to get used to working with the CAS. 

As for RQ~2, according to NASA-TLX, the perceived workload was reduced by an average of 26.8~\% with the CAS. The paired t-test showed a significant large effect for the RTLX difference between the CAS and the paper-based workflow.
Feedback from participants indicated that searching for documents in the paper file was a major problem that could be eliminated through the use of the CAS. Furthermore, participants stated that evaluating the wall thickness measurement table (Figs.~\ref{fig:paper_action4_points} and \ref{fig:paper_action4_table}) is very challenging, whereas the CAS was easy to use in this regard. This might be the reason action 4 was often skipped in the paper-based workflow.

All three AR functions were rated by the participants on average between important and very important. However, we observed that some participants mostly ignored the screen-based AR during the task and focused on the touchscreen with the 3D model instead. During the feedback, they stated that the AR technology was very new to them. Looking at the turbine blade's 3D model on the touchscreen monitor was more familiar to them. The AR assistance functions might be more useful during a real grinding process. In the present user study, participants could only indicate the physical repair process.

\section{Conclusion}
All ten participants preferred the CAS over their current paper-based workflow. The UMUX-LITE ratings showed better perceived usefulness and perceived ease of use for the CAS. Additionally, all assistance functions were rated on average between important and very important. 

In conclusion, the user study showed that the developed CAS can reduce the high-cost manual repair time for turbine blades of experienced metalworkers. Additionally, the digital work instructions with AR reduced the perceived workload. Furthermore, we could observe that the participants skipped inconvenient work steps with the paper file. With the assistance system, on the other hand, all the required information was conveniently displayed by default.

In future work, we would like to measure long-term effects by repeating the study after the workers had more time to get used to working with the assistance system. Additionally, we would like to test if the assistance system has a positive effect on the learning curve and the number of errors of new employees. We hypothesize that new employees who are not yet familiar with the paper file might learn the turbine blade repair process faster with the assistance system than without it. Finally, we would like to take a closer look at the advantages of the AR display compared to the digital work instructions shown on the touchscreen. 

\section*{Declaration of Competing Interest}
The authors declare that they have no known competing financial interests or personal relationships that could have appeared to influence the work reported in this paper.

\section*{Acknowledgements}
This work was part of the project MRO 2.0 - Maintenance, Repair, and Overhaul (ProFIT-10167454) and was supported in part by the European Regional Development Fund (ERDF). 

We would like to thank our project partners from Siemens Energy, Gestalt~Robotics, YOUSE, and Fraunhofer Institute for Production Systems and Design Technology for their work in the development process of the assistance system.

\addtolength{\textheight}{-2cm}   % This command serves to balance the column lengths
                                  % on the last page of the document manually. It shortens
                                  % the textheight of the last page by a suitable amount.
                                  % This command does not take effect until the next page
                                  % so it should come on the page before the last. Make
                                  % sure that you do not shorten the textheight too much.

%%%%%%%%%%%%%%%%%%%%%%%%%%%%%%%%%%%%%%%%%%%%%%%%%%%%%%%%%%%%%%%%%%%%%%%%%%%%%%%%

%%%%%%%%%%%%%%%%%%%%%%%%%%%%%%%%%%%%%%%%%%%%%%%%%%%%%%%%%%%%%%%%%%%%%%%%%%%%%%%%

%%%%%%%%%%%%%%%%%%%%%%%%%%%%%%%%%%%%%%%%%%%%%%%%%%%%%%%%%%%%%%%%%%%%%%%%%%%%%%%%

%%%%%%%%%%%%%%%%%%%%%%%%%%%%%%%%%%%%%%%%%%%%%%%%%%%%%%%%%%%%%%%%%%%%%%%%%%%%%%%%

\bibliographystyle{IEEEtran}

\bibliography{paper}

\appendix
\begin{figure*}[hbt]
     \centering
     \begin{subfigure}[b]{0.32\textwidth}
         \centering
         \includegraphics[height=\textwidth]{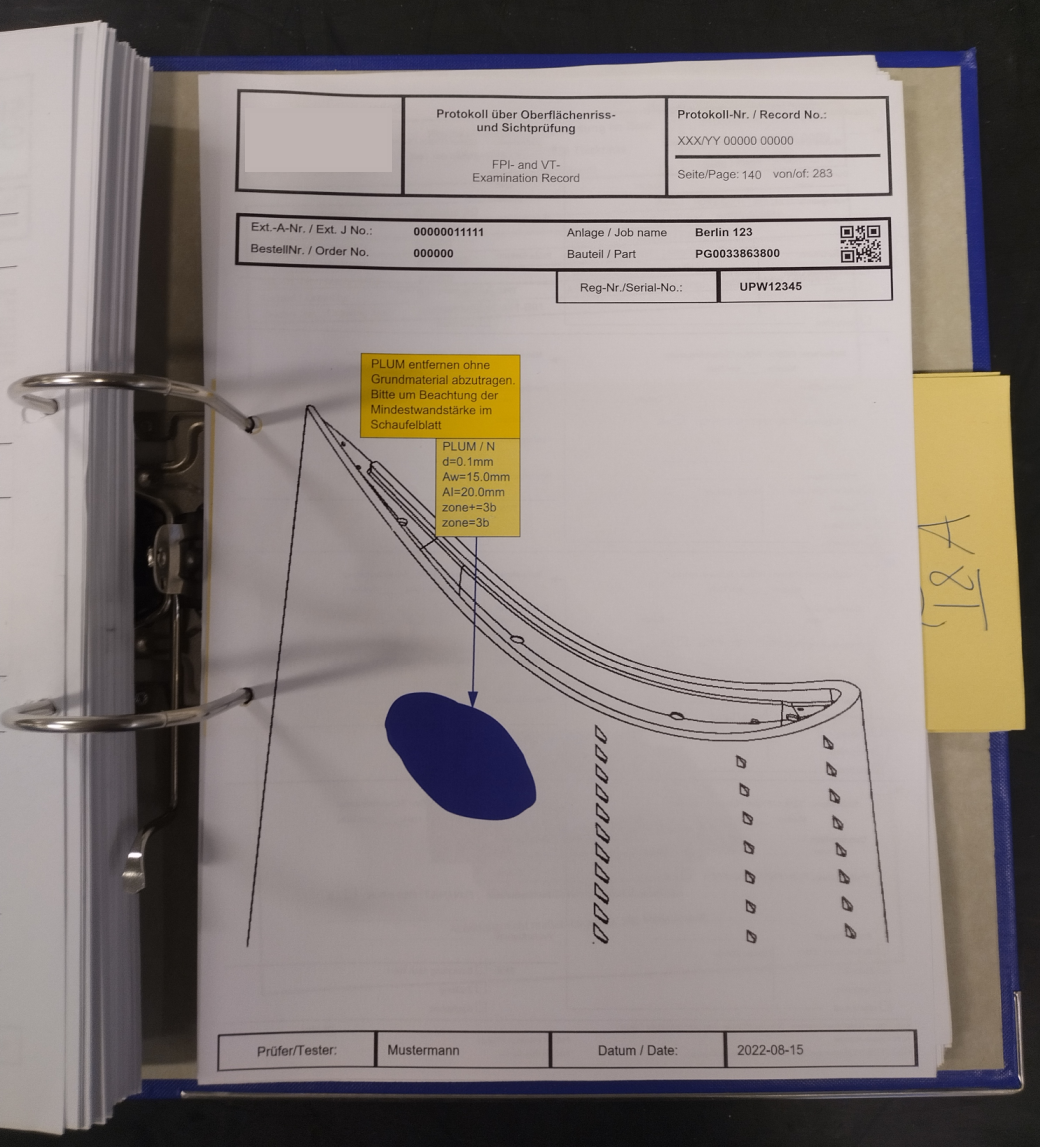}
         \caption{Defect information for one specific turbine blade (action~2)}
         %\label{fig:GUI_OVERVIEW}
     \end{subfigure}
     \hfill
          \begin{subfigure}[b]{0.32\textwidth}
         \centering
         \includegraphics[height=\textwidth]{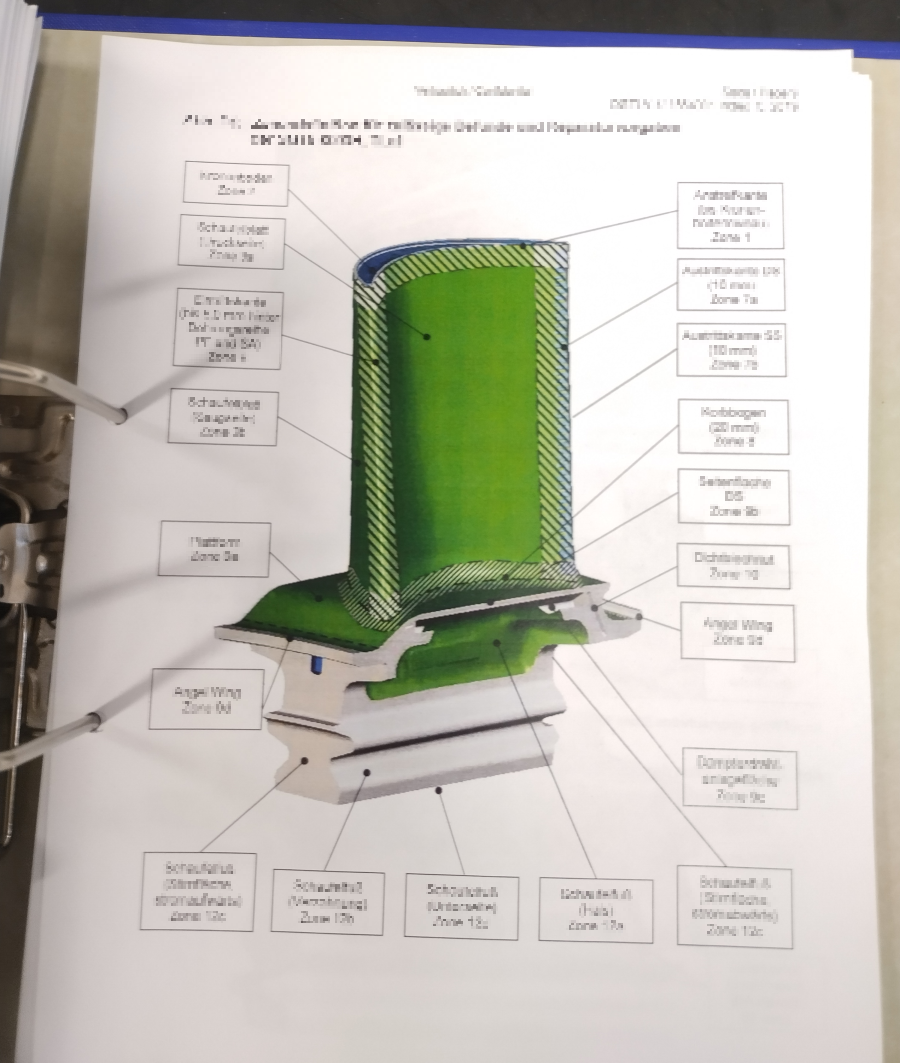}
         \caption{Zoning documentation for the product family (action~3)}
         \label{fig:paper_action3}
     \end{subfigure}
     \hfill
     \begin{subfigure}[b]{0.32\textwidth}
         \centering
         \includegraphics[height=\textwidth]{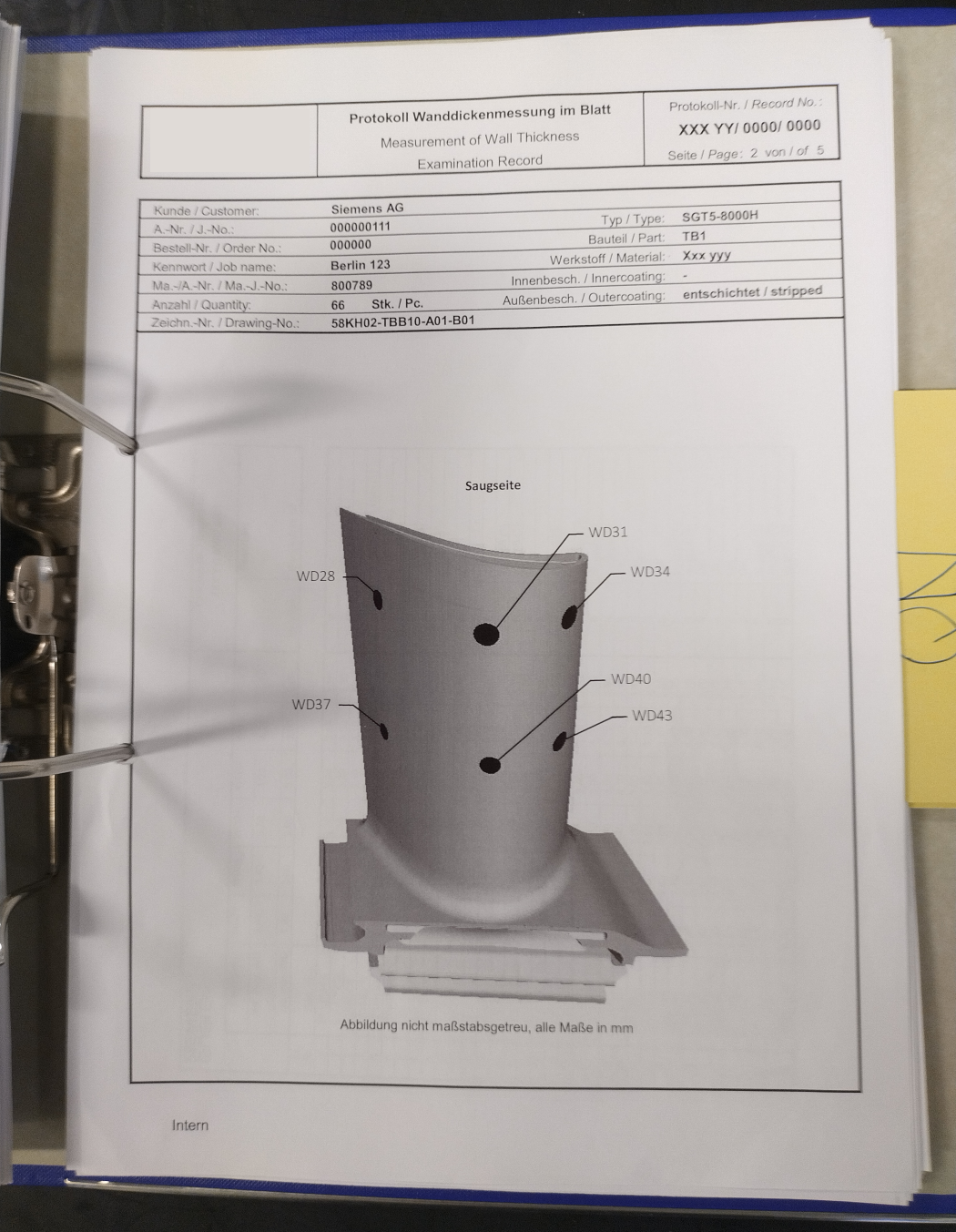}
         \caption{Wall thickness measurement spots for the product family (action~4)}
         \label{fig:paper_action4_points}
     \end{subfigure}\\
    \vspace{1cm}
     \begin{subfigure}[b]{0.49\textwidth}
         \centering
         \includegraphics[height=0.6\textwidth]{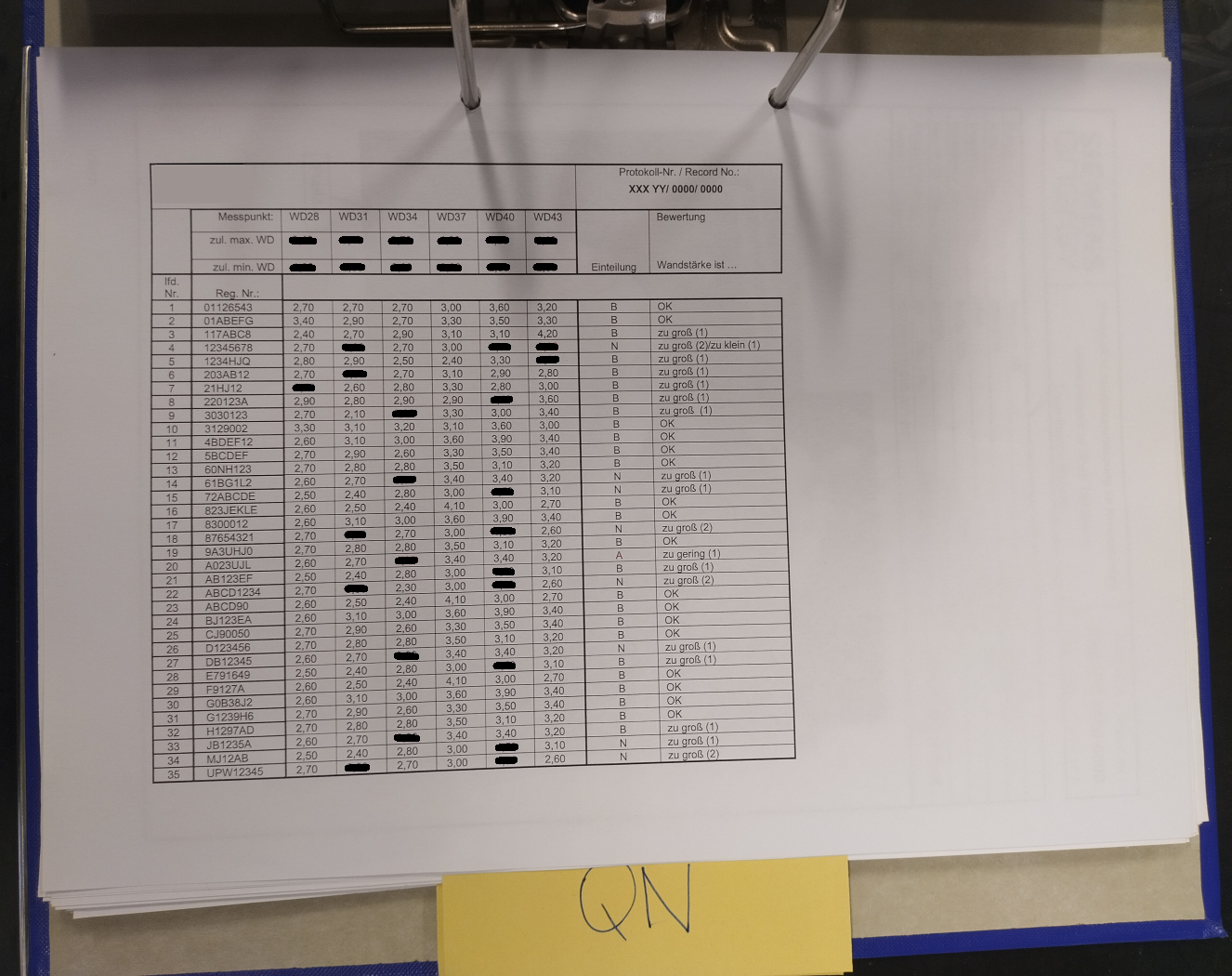}
         \caption{Wall thickness measurements for multiple serial numbers~(action~4)}
         \label{fig:paper_action4_table}
     \end{subfigure}
     \hfill
     \begin{subfigure}[b]{0.49\textwidth}
         \centering
         \includegraphics[height=0.6\textwidth]{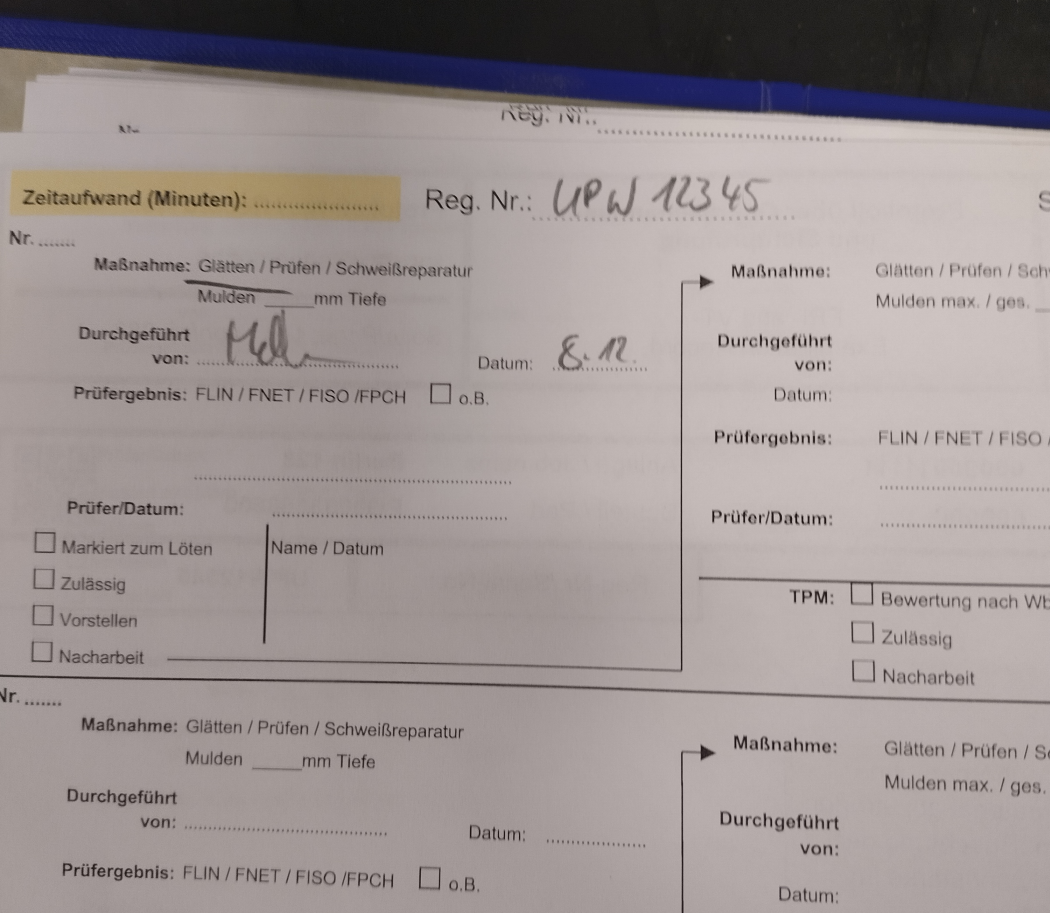}
         \caption{User input to document finished repair work for one turbine blade (action~6)}
         %\label{fig:PAPER_DOCUMENTATION}
     \end{subfigure}
        \caption{Repairing a turbine blade with the established paper file}
        \label{img:paper_appendix}
\end{figure*}

\begin{figure*}[hbt]
     \centering
     \begin{subfigure}[b]{0.32\textwidth}
         \centering
         \includegraphics[width=\textwidth]{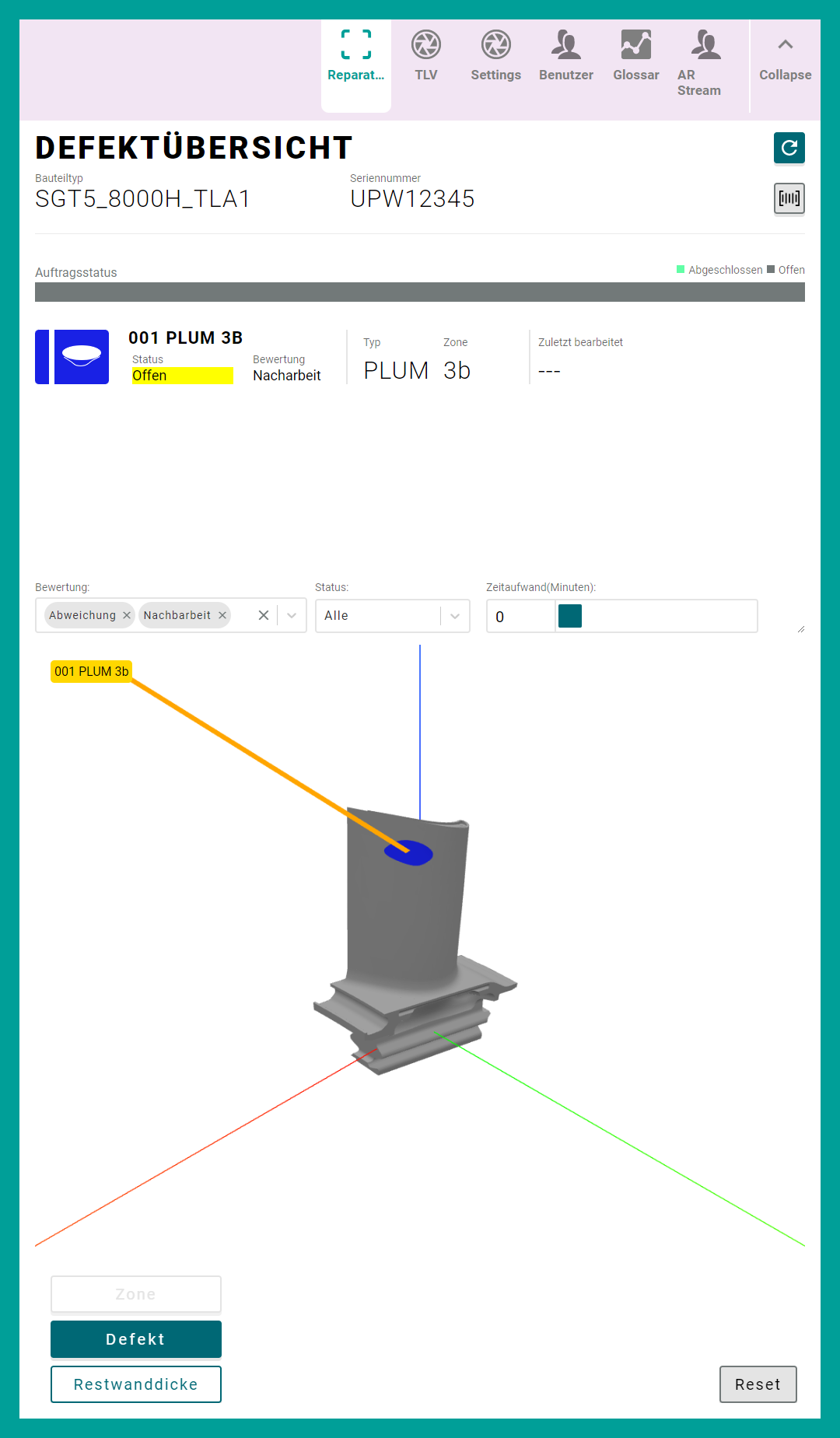}
         \caption{Overview of all defects for one turbine blade (action 2)}
         \label{fig:DWI1}
     \end{subfigure}
     \hfill
     \begin{subfigure}[b]{0.32\textwidth}
         \centering
         \includegraphics[width=\textwidth]{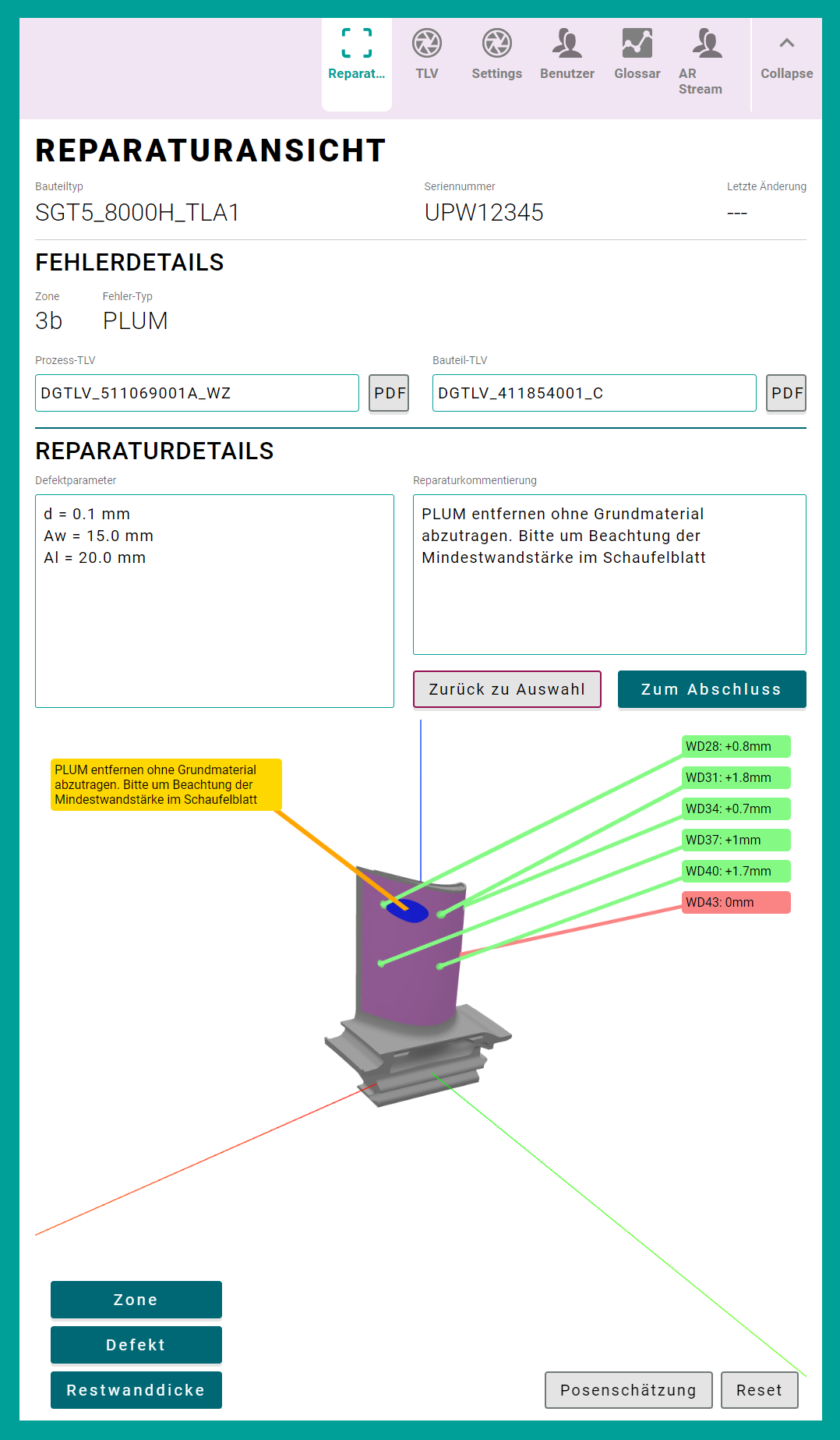}
         \caption{Digital work instructions for the selected defect (actions 2, 3, 4)}
         \label{fig:DWI2}
     \end{subfigure}
     \hfill
     \begin{subfigure}[b]{0.32\textwidth}
         \centering
         \includegraphics[width=\textwidth]{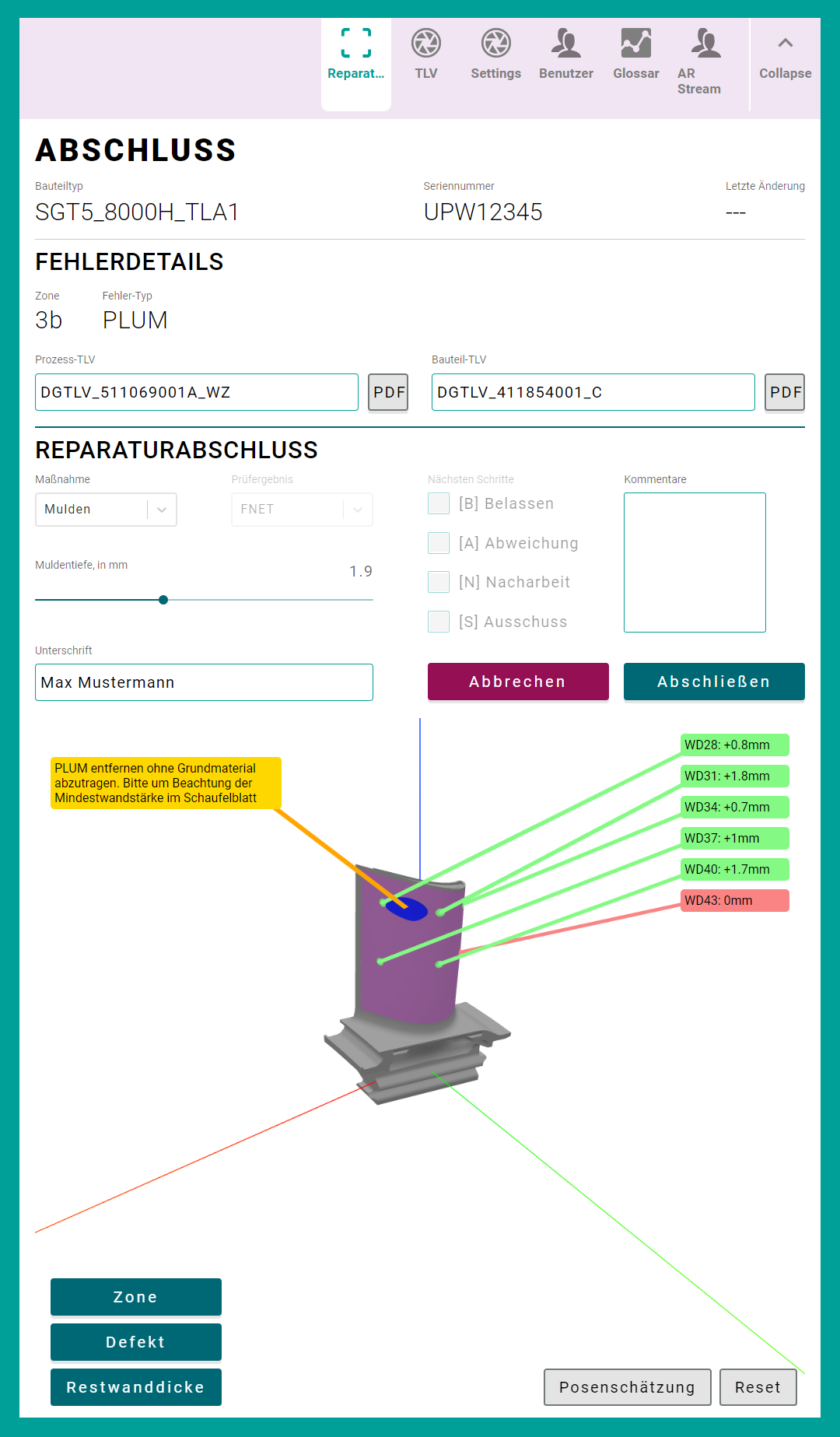}
         \caption{User input to document performed repair work (action 6)}
         \label{fig:DWI3}
     \end{subfigure}\\
    \vspace{1cm}
     \begin{subfigure}[b]{0.99\textwidth}
        \centering
         \includegraphics[width=0.7\textwidth]{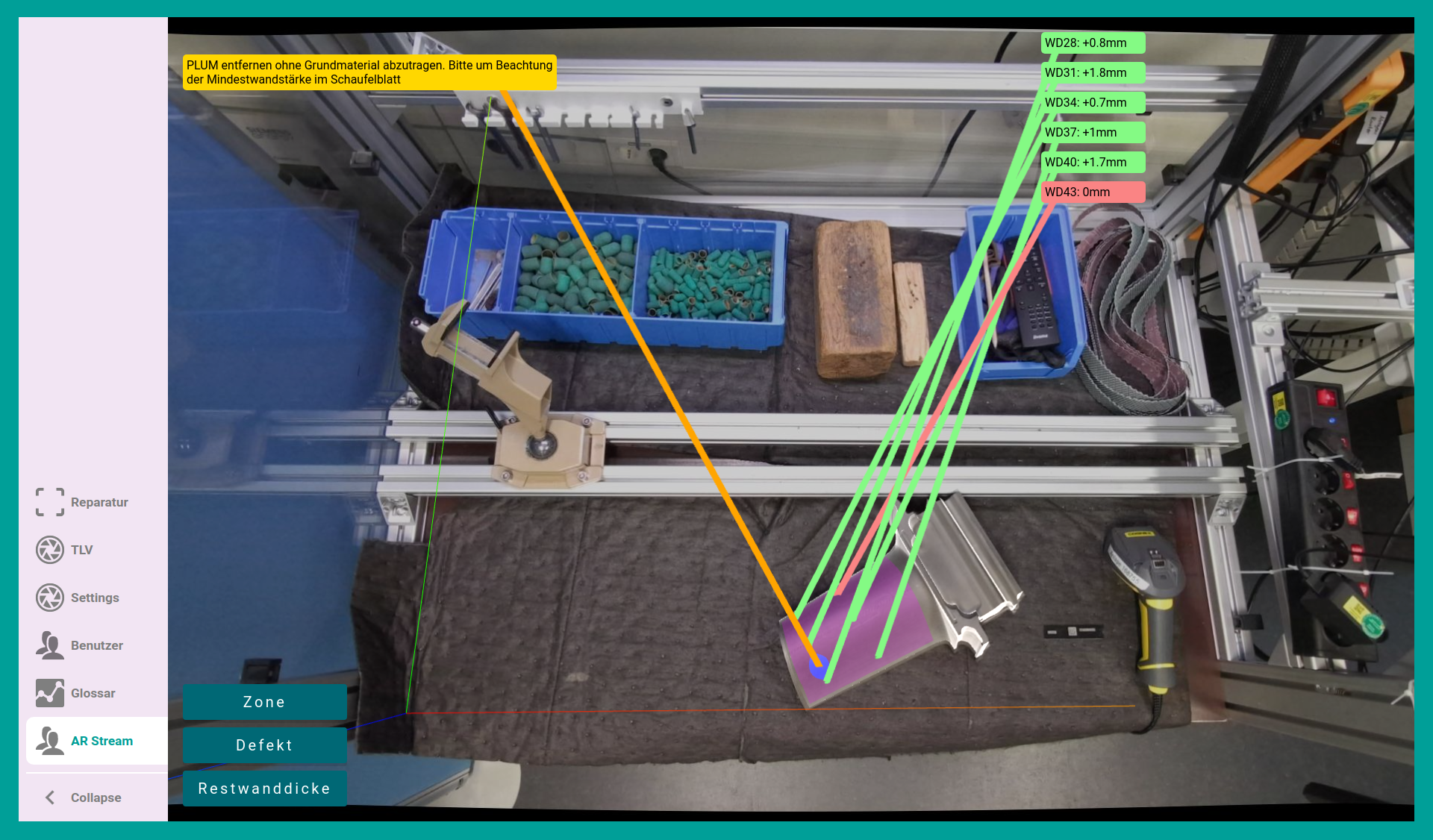}
         \caption{AR display with spatially registered work instructions (action 2, 3, 4)}
         \label{fig:AR}
     \end{subfigure}
        \caption{Repairing a turbine blade with the CAS}
        \label{img:CAS_appendix}
\end{figure*}

\end{document}